\documentstyle[prl,floats,aps,twocolumn,epsf]{revtex}
\tighten
\begin{document}
\draft
\twocolumn[\hsize\textwidth\columnwidth\hsize\csname
@twocolumnfalse\endcsname


\title{Late-time decay of scalar perturbations outside rotating
       black holes}
\author{Leor Barack \and Amos Ori}
\address {Department of Physics,
          Technion---Israel Institute of Technology, Haifa, 32000, Israel}
\date{\today}
\maketitle

\begin{abstract}
We present an analytic method for calculating the late-time tails
of a linear scalar field outside a Kerr black hole.
We give the asymptotic behavior at timelike infinity (for
fixed $r$), at future null infinity, and along the event horizon (EH).
In all three asymptotic regions we find a power-law decay.
We show that the power indices describing the decay of
the various modes at fixed $r$ differ
from the corresponding Schwarzschild values.
Also, the scalar field oscillates along the null generators of the EH
(with advanced-time frequency proportional to the mode's magnetic
number $m$).
\end{abstract}


\vspace{2ex}
]

The {\em no hair} principle for black holes implies that
gravitational field outside a generic black hole
relaxes at late time to the stationary Kerr-Newman geometry.
For linear test fields (either scalar, electromagnetic, or gravitational)
outside a spherically-symmetric Schwarzschild black hole,
it was shown by Price \cite{Price72} that all
radiative multipoles die off
at late time with a $t^{-2l-3}$ power-law tail \cite{nostatic},
where $l$ is the mode's multipole number, and $t$ is the
Schwarzschild time coordinate.
Later, this result was confirmed
using several different techniques, both analytic
and numerical \cite{Gundlach94I,Leaver86,Ching95,BarackII}.
The relevance of the perturbative (linear) results
to the fully nonlinear late-time behavior was demonstrated
in numerical simulations of a
fully non-linear, self-gravitating, spherically-symmetric scalar field
\cite{Gundlach94II,Burko97}.

It is well known, however, that realistic astrophysical
black holes are spinning and not spherically-symmetric \cite {bardeen}.
Therefore, for astrophysical applications
it is extremely important to generalize the above analyses from the
Schwarzschild background to the more realistic Kerr background.
A first progress in this direction has been achieved recently with the
numerical simulation of linear
fields in the Kerr background, by Krivan {\em et al.}
\cite{Krivan96,Krivan97}.
Yet, a thorough analytic treatment of this problem has not been
carried out so far \cite{ori,Hod1}.

The goal of this Letter is to present an analytic method for
calculating the late time behavior of a linear massless scalar field
outside a Kerr black hole. This method was recently applied to
the simpler Schwarzschild case as a test-bed
\cite{BarackII}, in which case the well known
late-time inverse-power tails were reproduced.
In this Letter we outline the application of this method to
the Kerr case, and present the main results.
In particular, we calculate the power indices characterizing the late-time
decay of the various modes at future null infinity, at fixed $r$, and
at the EH.
Quite interestingly, we find that at fixed $r$ these indices are
different than those found in the Schwarzschild case.
Full details of the calculations are given
in Ref.\ \cite {detail}. Throughout this paper we shall use the standard
Boyer-Lindquist coordinates $t,r,\theta,\varphi$, and relativistic units
$C=G=1$.

The main difficulty in analyzing perturbations over a Kerr
background is the nontrivial dependence on $\theta$. The
separation of variables by the Teukolsky equation \cite
{Teukolsky72} is only applicable to the Fourier-decomposed field,
because the spheroidal harmonics used for the separation of the
$\theta$, $\varphi$ variables explicitly depend on the temporal
frequency $\omega$. The final goal is to calculate the late-time
decay of the field, along with its angular dependence, in terms of
the time $t$. Obviously, an expression of this angular dependence
in terms of the ($\omega$-dependent) spheroidal harmonics would be
useless. This motivates one to carry out the entire analysis in
terms of the {\it spherical} harmonics $Y_{l}^{m}$.
The difficulty is, however, that
due to the breakdown of spherical symmetry, modes of different $l$
(but the same $m$) are coupled; Namely, there are "interactions"
between modes. The main challenge is to handle this interaction
and to analyze its effect on the late-time decay.

In principle, it is possible to carry out the analysis in the
frequency domain, and then Fourier-integrate over all frequencies
to recover the late-time behavior in the time domain, as was done
in the Schwarzschild case \cite{Leaver86,Ching95}. We find it very
difficult, however, to properly apply this method to the present
problem, due to the following reason. In the Schwarzschild case
the analysis can be much simplified by taking the limit $\omega
\to 0$. In the Kerr case, this limit does not lead to a correct
description of the interaction between modes \cite{Hod1}. As we
show below, it is this interaction which dominates the late-time
decay of the modes $l\ge 2$ at fixed $r$. In the limit $\omega \to
0$ one simply (and incorrectly) recovers the Schwarzschild result
\cite{Hod2}.

In view of these considerations,
we found it simpler to carry out the entire analysis
in the time domain (i.e. without a Fourier decomposition).
To overcome the difficulties caused by the interaction between modes,
in the first part of the analysis we use an iteration scheme
in which we iterate over the interaction term
(along with the other curvature-induced terms in the field equation).
In the second part of the analysis we use the
{\it late-time expansion}, which
is essentially an expansion in inverse powers of advanced time.
(Both methods are generalizations of those used
in Refs.\ \cite{BarackI,BarackII} for the spherical case).

The Klein-Gordon field $\Phi$ in Kerr geometry obeys
\begin{eqnarray} \label{eq1}
\lefteqn{
\left[\frac{(r^{2}+a^{2})^2}{\Delta}-a^{2}\sin^{2}\theta\right]
\Phi_{,tt} - \left(\Delta \Phi_{, r}\right)_{,r} +
\frac{4Mar}{\Delta} \Phi_{,t\varphi} }                 \nonumber\\
 &&+  \left(\frac{a^{2}}{\Delta}-\frac{1}{\sin^{2}\theta}\right)
\Phi_{,\varphi\varphi} -
\frac{1}{\sin\theta}\left(\Phi_{,\theta}\sin\theta\right)_{,\theta}
= 0,
\end{eqnarray}
where $M$ and $a$ are, correspondingly,
the black hole's mass and specific angular momentum, and
$\Delta\equiv r^{2}-2Mr+a^{2}$.
Decomposing $\Phi$ into spherical harmonics in the standard manner,
$\Phi=\sum_{lm} \phi^{lm}(t,r) \, Y_{l}^{m}(\theta,\varphi)$,
and defining
$\Psi^{lm}\equiv\sqrt{r^{2}+a^{2}}\;\phi^{lm}$,
the original field equation (\ref {eq1}) becomes (for each $m$)
\begin{eqnarray} \label{eq4}
\lefteqn{\Psi_{,uv}^{l}+V_{K}^{lm}(r)\Psi^{l}+
i\frac{mMar}{(r^{2}+a^{2})^{2}}\Psi_{,t}^{l} }           \nonumber\\
&&+\frac{a^{2}\Delta}{(r^{2}+a^{2})^{2}}
\left[c_{0}\Psi_{,tt}^{l}+c_{-}\Psi_{,tt}^{l-2}+c_{+}\Psi_{,tt}^{l+2}
\right] = 0 ,
\end{eqnarray}
where $c_{0}$, $c_{-}$ and $c_{+}$ are certain coefficients
depending on $l$ and $m$ (with $c_{-}^{m=l,l-1}=0$; no other
coefficients vanish), and where
\begin{eqnarray} \label{eq6}
4V_{K}^{lm}&=&\Delta(r^2+a^2)^{-4}\left(2Mr^3+a^2r^2-4Ma^2r+a^4\right)
\nonumber\\
&&-(r^2+a^2)^{-2}\left[m^2a^2-l(l+1)\Delta\right].
\end{eqnarray}
The coordinates $u$ and
$v$ are defined by $u=t-r_{*}$ and $v=t+r_{*}$, with $r_{*}(r)$
obeying $dr_{*}/dr=(a^{2}+r^{2})/\Delta$. [In Eq.\ (\ref {eq4}),
and also in most of the equations below, we omit the index $m$ for
brevity. Note that due to the axial symmetry, modes with different
$m$ do not interact.]

Equation~(\ref{eq4}) is an infinite set of coupled equations for
the various modes of the field, with the last two terms in the square
brackets describing the interactions between modes of different $l$.

The set-up of initial data for the evolution problem is similar to
that used in Ref.\ \cite{BarackII} for the Schwarzschild case
[see figure 2 and Eq.\ (7) therein].
That is, we assume that $\Phi$ is specified along a pair of hypersurfaces
$v=v_0$ and $u=u_0$, and without loss of generality we take $v_0=0$.
For convenience we consider a situation of an outgoing pulse at $v=0$,
which starts immediately after the outgoing ray $u=u_0$.
We further assume
that $-u_0\gg M$, and that the pulse is
arbitrarily-shaped but relatively short,
which considerably simplifies the analysis \cite{BarackI,BarackII}.
Although this type of initial data is not the most general one,
it is nevertheless reasonably generic, and we expect the resultant
asymptotic behavior to be characteristic of the generic situation.
We also assume here that the initial outgoing pulse
has a rather generic angular distribution, so it includes all the
spherical harmonics (and especially the component $l=0$).
(Later we shall briefly discuss the more subtle case,
in which no $l=0$ component is present
at the initial data.)

To evolve these initial data and analyze the late-time behavior,
we shall proceed in two
steps. In the first step, we calculate the late time
(i.e.\ $u\gg M$) form of the field at future null-infinity
($v\rightarrow \infty$). In the second step we derive an
expression for the field at a fixed $r$ at $t\gg M$
(and also along the EH at $v\gg M$).

\paragraph* {Asymptotic behavior at future null-infinity.}

We now apply the iteration scheme introduced in
Refs.\ \cite{BarackI,BarackII}, and decompose $\Psi^{lm}$ as
\begin{equation} \label{eq7}
\Psi^{lm}=\sum_{N=0}^{\infty}\Psi_{N}^{lm}.
\end{equation}
The components $\Psi_{N}^{lm}$ are defined by the hierarchy of equations
\begin{equation} \label{eq8}
        (\Psi_{N,}^{l})_{,uv}+V_{0}^{l}\Psi_{N}^{l}=S_{N}^{l},
\end{equation}
where $S_{0}^{l}\equiv 0$ and (for $N\geq 1$)
\begin{eqnarray} \label{eq9}
\lefteqn{S_{N}^{l}        \equiv
         -(\delta V_{K}^{l})\Psi_{N-1}^{l}
         -i\frac{mMar}{(r^{2}+a^{2})^{2}}(\Psi_{N-1}^{l})_{,t}}
                                                              \nonumber\\
&& -\frac{a^{2}\Delta}{(r^{2}+a^{2})^{2}}
   \left[    c_{0}\Psi^{l}_{N-1}+c_{+}\Psi^{l+2}_{N-1}
   +c_{-}\Psi^{l-2}_{N-1}                \right]_{,tt},
\end{eqnarray}
along with the initial conditions $\Psi^{l}_{0}=\Psi^{l}$
and $\Psi_{N\geq 1}^l=0$ at $v=0$ and $u=u_0$.
Here, $V_{0}^l(r_{*})$ is the Minkowski-like potential defined in
Ref.\ \cite{BarackII} as a function of $r_{*}$
[Eqs.\ (8,60) therein], and $\delta V_{K}^l(r)\equiv V_{K}^l-V_{0}^l$.
Formal summation over $N$ recovers the original field equation and
initial data for the complete fields $\Psi^l$.

The field equation (\ref{eq8}), together with the above initial
conditions, constitutes a hierarchy of initial-value problems for the
various functions $\Psi_{N}^{lm}$, which, in principle, we may solve
one by one (first for $N=0$, then for $N=1$, etc.).
Notice that the potential $V_{0}(r_{*})$ (and
hence the entire $N=0$ equation) is independent of
the spin parameter $a$.
The solution of this equation, the function $\Psi_{0}^{l}$,
is given explicitly in Ref.\ \cite{BarackII}
(see section IV therein). This function
decays exponentially at late time, so it does not contribute to
the power-law tail. Rather, it serves as a source for
terms $\Psi_{N\geq 1}$, which do provide power-law tails.
For each $N\geq 1$, the field equation can be formally solved in terms
of a Green's function:
\begin{equation} \label{eq11}
\Psi_{N}^{l}(u,v)=\int_{u_{0}}^{u}\!\!\!du'\int_{0}^{v}\!\!\!dv'
G^l(u,v;u',v')S_N^l(u',v'),
\end{equation}
where $G^l(u,v;u',v')$ is the (retarded) Green's function
associated with the zero-order operator
$\partial_{u}\partial_{v}+V^l_{0}$. An analytic expression for $G$
was derived in section V of Ref.\ \cite{BarackII}.
This, in principle, enables the solution of the field equation
(\ref{eq8}) for all $N$ and $l$.

The functions $\Psi_{1}^{l}$ will primarily concern us here,
because it is the term $N=1$ which dominates the late-time
behavior of $\Psi^{l}$ at null infinity
in the generic situation. It is convenient to
consider separately the contributions from the various terms in
Eq.\ (\ref {eq9}) [through Eq.\ (\ref{eq11})] to $\Psi_{1}^{l}$.
Consider first the contribution from the term proportional to
$\delta V_{K}^{l}$. This potential can be expressed as
$\delta V_{K}^{l}=\delta V_{S}^{l}+\delta V_{a}^{l}$,
where $\delta V_{S}^{l}$ is the
($a$-independent) corresponding Schwarzschild
contribution, and $\delta V_{a}^{l}$ is an $a$-dependent correction term.
A direct calculation shows that at $r\gg M$, $\delta V_{a}^{l}$ decays
faster than $\delta V_{S}^{l}$ by a
factor proportional to $a^2/(Mr_*)$. An explicit
evaluation of the integral in Eq.\ (\ref{eq11})
then yields that this extra
factor leads to an extra $u^{-1}$ factor in the late-time asymptotic
behavior of $\Psi_{1}^{l}$ at null infinity \cite{detail}.
Thus, the dominant contribution to $\Psi_{1}^{l}$
from $\delta V_{K}^{l}$ at null infinity,
which we denote by $\hat{\Psi}_1^l$, is the same as
in the Schwarzschild case (cf. Eq.\ (58) in Ref.\ \cite{BarackII}):
\begin{equation} \label{eq12}
\hat{\Psi}_{1}^{l}(u\gg M)\cong A_{l}\,u^{-l-2},
\end{equation}
where $A_{l}$ is given explicitly in \cite{BarackII} as a linear
functional of the $l$-component of the initial pulse .

Consider next the contribution to $\Psi_{1}^{l}$ due to the other
part of $S_1^l$, i.e.\ the part containing $t$-derivatives in
Eq.\ (\ref{eq9}).
A direct evaluation of the integrals in Eq.\ (\ref{eq11}) shows that
the late-time contribution of this part at null infinity is proportional
to $u^{-l-3}$ or smaller \cite{detail},
and is hence negligible. The only exception is the
contribution from $\Psi^{l-2}_{0,tt}$ (for $l\geq 2$), which is
proportional to $u^{-l-2}$ too, so it does not cause a qualitative change
in the asymptotic decay (\ref {eq12}). Moreover, the coefficient of this
term is reduced by a factor proportional to $(a/u_0)^2\ll 1$ compared to
$A_l$, so the overall tail amplitudes
are still given by Eq.\ (\ref {eq12}) at the leading order.

We still need to consider the terms $N\geq 2$.
A complete analysis of these terms has not been carried out yet.
In the Schwarzschild case, simple considerations suggest
that at null infinity all these terms decay like $u^{-l-2}$, though with
coefficients smaller than that of $\Psi_{1}$ by a
factor $(M/u_0)^{N-1}$.
(This was also verified by numerical simulations \cite{BarackII},
which also suggested convergence of the sum (\ref{eq7}) at null infinity.)
Hence the $N\geq 2$ terms do not alter the power index, and, moreover,
in the case $-u_0\gg M$ considered here, they do not significantly affect
the coefficients. All these considerations
apply to the Kerr case as well \cite{detail,onlyforgeneric}.
Assuming that the terms $N\geq 2$ indeed behave in that manner,
we find that at late time, the scalar field at null infinity is
given by
\begin{equation} \label{eq13}
\begin{array}{ll}
\Psi^{l}\cong A_{l}\, u^{-l-2}
        &     (\text{null infinity, $u\gg M$}),
\end{array}
\end{equation}
where the coefficients $A_{l}$ coincide with
those of Eq.\ (\ref {eq12})
to the leading order in $M/u_0$.

\paragraph* {Derivation of $\Phi$ at $r$={\rm const}: the
late-time expansion.}

We now derive an expression for the late-time behavior
at any fixed $r$ outside the black hole and along the EH,
accurate to the leading order in $M/t$ or $M/v$, respectively.
To that end we employ the late-time expansion used in the
Schwarzschild case (cf.\ Ref.\ \cite{BarackII}):
\begin{equation} \label{eq14}
\phi^{lm}(r,v)=\sum_{k=0}^{\infty}F_{k}^{lm}(r)v^{-k_{0}-k}.
\end{equation}
As it turns out, this expansion is consistent with the field equation,
with the regularity condition at the EH, and with the form of the field at
null-infinity. The parameter $k_0>0$ is by definition $l$-independent,
and will later be determined through matching to null infinity.
(For $l>0$ some of the first terms in the sum (\ref{eq14}) vanish,
as will become apparent below.)

Substituting the expansion (\ref{eq14})
in the field equation (\ref{eq4}), and
collecting terms of the same power in $v$, the partial differential
equation becomes an infinite coupled set of {\it ordinary} equations
for the unknown functions $F_{k}^{l}(r)$,
\begin{eqnarray} \label{eq15}
\lefteqn{\left[\Delta (F_{k}^{l})^{\prime}\right]^{\prime}+
\left[a^{2} m^{2}/\Delta-l(l+1) \right] F_{k}^{l}=Z_k^l\equiv}
\nonumber\\ && 2(k_{0}+k-1)\!\left[
(r^{2}+a^{2})(F_{k-1}^{l})^{\prime}+\left(r\!-\!2imMar/\Delta\right)
F_{k-1}^{l}\right.
\nonumber\\ && \left. +2a^{2} (k_{0}+k-2)\left(
c_{0}F_{k-2}^{l}+c_{+}F_{k-2}^{l+2}+c_{-}F_{k-2}^{l-2} \right)
\right],
\end{eqnarray}
where a prime denotes $d/dr$ (and where $F_{k'<0}^l\equiv 0$).
Here, too, the source term $Z_k^l$ exhibits interactions with
modes $l'\ne l$. However, since $Z_k^l$ only depends on terms
$F_{k'}^{l'}$ with $k'<k$, the system (\ref{eq15}) is effectively
decoupled, as the various ordinary equations may be solved one at
a time. It is possible to formally write down the general solution
for any $F_k^l$ via the Green's-function method \cite{detail}.
Consider first the term $k=0$, which dominates the overall
late-time asymptotic behavior at fixed $r$.
The function $F_{k=0}^l$
satisfies a homogeneous equation (actually the stationary
field equation), whose general solution is given by
$F_{0}^l=a_{l}P_{l}^{-\gamma}(\rho)+b_{l}P_{l}^{+\gamma}(\rho)$.
Here, $a_{l}$ and $b_{l}$ are (yet) arbitrary constants,
$\rho\equiv (2r-r_{+}-r_{-})/(r_{+}-r_{-})$ where $r_{\pm}\equiv
M\pm \left(M^{2}-a^{2}\right)^{1/2}$, and $P_{l}^{\pm \gamma}$ are
the two complex-conjugated {\em associated Legendre functions of
the first kind} \cite{GradRyz} with an imaginary index
$\gamma=im\left[2a/(r_{+}-r_{-})\right]$. The functions
$P_{l}^{\pm \gamma}$ (which are special cases of the
Hypergeometric function) have the form
\begin{equation} \label{eq18}
P_{l}^{\pm \gamma}(\rho)={\cal P}_{l}^{\pm \gamma}(\rho)\times
\left[(\rho+1)/(\rho-1)\right]^{\pm \gamma /2},
\end{equation}
in which ${\cal P}_{l}^{\pm \gamma}$ are (complex) polynomials of
order $l$ (nonvanishing at $r\to r_+$).
For $m\neq 0$ these functions oscillate rapidly towards the EH
($r\rightarrow r_{+},\;\rho\rightarrow 1$),
\begin{equation} \label{eq19}
P_{l}^{\pm \gamma}(r\rightarrow r_{+})\propto
(\rho-1)^{\mp \gamma /2}
\propto\exp(\mp im\Omega_{+} r_{*}),
\end{equation}
where $\Omega_{+}\equiv a/(2Mr_{+})$.

One of the two coefficients $a_l,b_l$ is
to be determined from the regularity
condition at the EH. Here one must recall that
the Boyer-Lindquist coordinate $\varphi$ is singular
at the EH. Using the regularized azimuthal coordinate
$\tilde{\varphi}_{+}\equiv \varphi-\Omega_{+}t$
\cite{Chandra83} instead, one finds
\begin{equation} \label{eq19a}
e^{im\varphi}=[e^{im\tilde\varphi_{+}}\,e^{im\Omega_{+}v}]\;
               e^{-im\Omega_{+}r_*}.
\end{equation}
Since the factor in square brackets is regular at the EH
(but the next factor is not), it
follows from the regularity condition that $b_l=0$, hence
$F_{0}^l=a_{l}P_{l}^{-\gamma}(\rho)$.

The parameter $a_l$ is to be determined through matching to null infinity.
The asymptotic form of $F_0^l$ as $r\to \infty$ is
$F_{0}^l(r\gg M)\propto r^{l}$.
Substitution in Eq.\ (\ref {eq14})
(taking into account the contribution of the terms
$k>0$ as well, which are not negligible at null infinity,
as explained in \cite{BarackII}),
one obtains at null infinity
$\Psi^{l}\propto a_l u^{l+1-k_{0}}$ \cite{detail}.
Matching this expression to Eq.\ (\ref {eq13}) for $l=0$ yields
$k_0=3$. This value of $k_0$ yields a consistent matching for any
$l$, implying $a_{l\geq1}=0$
(that is, the modes $l\geq 1$ are excited only at $k>0$).
One finds that the dominant mode $l=0$ decays
like $v^{-3}\propto t^{-3}$ at fixed
$r$ (and large $t$), as in the Schwarzschild case.

The interaction between modes has a crucial effect on the decay rate
of modes $l\ge 2$.
Without this interaction, one can verify that a mode $l,m$
is excited at $k=2l$, leading to a decay rate $t^{-2l-3}$
(as in the Schwarzschild case)\cite{explain}.
The interaction changes this situation.
Consider, for example, the mode $l=2,m=0$.
This mode has a vanishing source term $Z_{k=1}^{l=2}$
(as $F_{k=0}^{l=2}\equiv 0$), and one can show (using the argument of
\cite{explain}) that $F_{k=1}^{l=2}(r)\equiv 0$.
On the other hand, at $k=2$ there is a non-vanishing source term
$Z_{k=2}^{l=2}\propto c_- F_{k=0}^{l=0}$, which
necessarily leads to a non-vanishing function $F_{k=2}^{l=2}(r)$.
Thus, the decay rate of this
mode at fixed $r$ is $v^{-k_0-k}=v^{-5}\propto t^{-5}$, which differs
from the corresponding Schwarzschild rate, $t^{-7}$.
This simple consideration
is easily extended to all other modes $m,l$, and one finds \cite{detail}
\begin{equation} \label{eq31}
\begin{array}{ll}
\Psi^{lm}\propto t^{-l-\left|m\right|-3-q}
         & (\text{fixed $r$,\ \ $t\gg M,|r_*|$}),
\end{array}
\end{equation}
where $q=0$ for even $l+m$ and $q=1$ for odd $l+m$.
The late-time behavior of a mode $l,m$
at the EH (expressed in regular coordinates) is found to be
\begin{equation} \label{eq32}
\Psi^{lm}Y_{l}^{m}(\theta,\varphi)\propto
Y_{l}^{m}(\theta,\tilde\varphi_+)\,
e^{im\Omega_{+}v}\; v^{-l-\left|m\right|-3-q}.
\end{equation}
(This power index may be changed if the relevant function $F_k^l(r)$
happens to vanish as $r\to r_+$.)

In summary, the late-time behavior of the various modes in the three
asymptotic regions is given in Eqs.\
(\ref{eq13}), (\ref{eq31}), and (\ref{eq32}).
(For the dominant mode $l=0$, the amplitude coefficients in all
three asymptotic regions are given explicitly in Ref.\ \cite{detail}.)
Our analysis indicates two phenomena special to the Kerr case:\\
A. {\em Oscillations along the EH} --- cf. Eq.\ (\ref{eq32}).
        \\
B. {\em The interaction between modes:}
        Due to this interaction, the power index
        at fixed $r$ is $l+\left|m\right|+3+q$.
        (When the interactions are ignored, one obtains the
        standard Schwarzschild exponent $2l+3$ \cite{Hod2}.)

Throughout this paper we have assumed that the initial pulse
includes all the modes (and in particular, the dominant mode
$l=0$). In non-generic situations in which the low-$l$ modes are
absent at the initial data, the interaction between modes may
dominate the overall late-time behavior already at null infinity.
For example, assume that
the initial pulse is a pure $l=2,m=0$ mode. Then, without the
interactions, at null infinity $\Psi$ would be dominated by
$\Psi_{N=1}^{l=2}\propto u^{-4}$. However, the interaction excites (at
$N=2$) an $l=0$ mode with a $u^{-2}$ tail, that
dominates the late-time behavior. This behavior has been
observed numerically by Krivan {\em et al.} \cite{Krivan96}.
In Ref.\ \cite{detail} this phenomenon will be discussed in more detail,
along with its implications to the asymptotic behavior at fixed $r$.

We should emphasize that despite the relative simplicity
of the calculation scheme
presented here, the mathematical question of convergence
of the various expansions involved is still open
(though there is evidence for convergence).
This is the situation even in the Schwarzschild case.
Additional mathematical subtleties arise in the Kerr case, which
we further discuss in \cite{detail}.

We finally note that numerical simulations \cite{Krivan96}
are consistent with our analytic results for
the power indices of the overall perturbation at fixed $r$
\cite{tildephi}. It will be interesting to numerically test our
prediction of the power index $l+\left|m\right|+3+q$ for the
individual $l,m$ modes at fixed $r$.

Note added: After this manuscript has been submitted, Hod analyzed
the mode coupling in Kerr spacetime \cite{Hod3}. In the scalar
field case he recovers our results.




\begin{references}
\bibitem{nostatic} For brevity, we shall
only consider in this Letter modes without initial static multipole.
\bibitem{Hod2} See Eq.\ (36) in \cite{Hod1},
which yields the power index $2l+3$.
\bibitem{onlyforgeneric}
This holds as long as the initial pulse
has a generic angular distribution, as we assume here.
This situation changes if
the small-$l$ initial modes are absent, as we discuss below.
\bibitem{explain}
Without interactions,
for any $k\leq 2l$, $F_{k'<k}^{l}=0$ yields $Z_k^l=0$, and the general
EH-regular homogeneous solution of Eq.\ (\ref{eq15}) is
$F_k^l=a_{lk}P_l^{-\gamma}(\rho)$, which in turn yields at null
infinity $\Psi^l\propto a_{lk}u^{l+1-(k_0+k)}$.
Matching to Eq.\ (\ref{eq13}) now yields $a_{lk}=0$ for any
$k<2l$, and $a_{lk}\ne 0$ for $k=2l$.
\bibitem{tildephi} The oscillations along the
event horizon are not manifested in
Ref.\ \cite{Krivan96}, because the Kerr coordinate
$\tilde{\varphi}$ is used there instead of our
coordinate $\tilde{\varphi}_{+}$. Both coordinates
are regular at the EH; however, the horizon's null generators
are lines of constant $\tilde{\varphi}_{+}$
(but varying $\tilde{\varphi}$).
Note that the oscillation of the scalar field along the horizon's
null generators is a coordinate-independent phenomenon.
\end{references}
\end{document}